\documentstyle[epsf]{mn}

\newif\ifAMStwofonts

\ifoldfss
  \ifCUPmtlplainloaded \else
    \NewTextAlphabet{textbfit} {cmbxti10} {}
    \NewTextAlphabet{textbfss} {cmssbx10} {}
    \NewMathAlphabet{mathbfit} {cmbxti10} {} 
    \NewMathAlphabet{mathbfss} {cmssbx10} {} 
  \fi
  \ifAMStwofonts
    \ifCUPmtlplainloaded \else
      \NewSymbolFont{upmath} {eurm10}
      \NewSymbolFont{AMSa} {msam10}
      \NewMathSymbol{\upi}     {0}{upmath}{19}
      \NewMathSymbol{\umu}     {0}{upmath}{16}
      \NewMathSymbol{\upartial}{0}{upmath}{40}
      \NewMathSymbol{\leqslant}{3}{AMSa}{36}
      \NewMathSymbol{\geqslant}{3}{AMSa}{3E}

    \fi
  \fi
\fi 

\ifnfssone
  \newmathalphabet{\mathit}
  \addtoversion{normal}{\mathit}{cmr}{m}{it}
  \addtoversion{bold}{\mathit}{cmr}{bx}{it}
  \newmathalphabet{\mathbfit} 
  \addtoversion{normal}{\mathbfit}{cmr}{bx}{it}
  \addtoversion{bold}{\mathbfit}{cmr}{bx}{it}
  \newmathalphabet{\mathbfss} 
  \addtoversion{normal}{\mathbfss}{cmss}{bx}{n}
  \addtoversion{bold}{\mathbfss}{cmss}{bx}{n}
  \ifAMStwofonts
    \ifCUPmtlplainloaded \else
and
your
      \UseAMStwoboldmath
      \makeatletter
      \new@mathgroup\upmath@group
      \define@mathgroup\mv@normal\upmath@group{eur}{m}{n}
      \define@mathgroup\mv@bold\upmath@group{eur}{b}{n}
      \edef\UPM{\hexnumber\upmath@group}
      \new@mathgroup\amsa@group
      \define@mathgroup\mv@normal\amsa@group{msa}{m}{n}
      \define@mathgroup\mv@bold\amsa@group{msa}{m}{n}
      \edef\AMSa{\hexnumber\amsa@group}
      \makeatother
      \mathchardef\upi  "0\UPM19
      \mathchardef\umu  "0\UPM16
      \mathchardef\upartial  "0\UPM40
      \mathchardef\leqslant  "3\AMSa36
      \mathchardef\geqslant  "3\AMSa3E
    \fi
  \fi
\fi 

\ifnfsstwo

  \def\spose#1{\hbox to 0pt{#1\hss}}
  \def\lta{\mathrel{\spose{\lower 3pt\hbox{$\mathchar"218$}}
     \raise 2.0pt\hbox{$\mathchar"13C$}}}
  \def\gta{\mathrel{\spose{\lower 3pt\hbox{$\mathchar"218$}}
     \raise 2.0pt\hbox{$\mathchar"13E$}}}
  \DeclareMathAlphabet{\mathbfit}{OT1}{cmr}{bx}{it}
  \SetMathAlphabet\mathbfit{bold}{OT1}{cmr}{bx}{it}
  \DeclareMathAlphabet{\mathbfss}{OT1}{cmss}{bx}{n}
  \SetMathAlphabet\mathbfss{bold}{OT1}{cmss}{bx}{n}
  \ifAMStwofonts
    \ifCUPmtlplainloaded \else
      \DeclareSymbolFont{UPM}{U}{eur}{m}{n}
      \SetSymbolFont{UPM}{bold}{U}{eur}{b}{n}
      \DeclareSymbolFont{AMSa}{U}{msa}{m}{n}
      \DeclareMathSymbol{\upi}{0}{UPM}{"19}
      \DeclareMathSymbol{\umu}{0}{UPM}{"16}
      \DeclareMathSymbol{\upartial}{0}{UPM}{"40}
      \DeclareMathSymbol{\leqslant}{3}{AMSa}{"36}
      \DeclareMathSymbol{\geqslant}{3}{AMSa}{"3E}
    \fi
  \fi
\fi 

\ifCUPmtlplainloaded \else
  \ifAMStwofonts \else 
    \def\upi{\pi}
    \def\umu{\mu}
    \def\upartial{\partial}
\fi

\title{Hipparcos open clusters and stellar evolution}
 
\author[V. Castellani et al.]
       {V.Castellani$^{1,2}$, S. Degl'Innocenti$^{3,4}$, P. G. Prada Moroni$^{3,4}$, V. Tordiglione$^{3}$\\
  $^1$ Osservatorio Astronomico di Roma, via Frascati 33, 00040 Monte Porzio Catone, Italy\\
  $^2$ INFN Sezione di Ferrara, via Paradiso 12, 44100 Ferrara, Italy \\
  $^3$ Dipartimento di Fisica, Universit\`a di Pisa, via Buonarroti 2, 56127 Pisa, Italy \\
  $^4$ INFN Sezione di Pisa, via Livornese 1291, 56010 S. Piero a Grado, Pisa, Italy \\
}

\pagerange{\pageref{firstpage}--\pageref{lastpage}}
\pubyear{ 00}

\begin{document}

\maketitle

\label{firstpage}

\begin{abstract}

By relying on recently improved  Hipparcos parallaxes for the
Hyades, Pleiades and Ursa Major clusters we find that stellar
models with updated physical inputs  nicely reproduce the location
in the color magnitude diagram of  main sequence stars of different
metallicities. Stars in the helium burning phase are also discussed,
showing that the luminosity of giants in the Hyades, Praesepe and Ursa Major clusters
appears to be in reasonable agreement with theoretical predictions. A short
discussion concerning the current evolutionary scenarios closes the paper.

\end{abstract}

\begin{keywords}
open clusters and associations:individual:Hyades, Pleiades, Ursa Major, Praesepe, stars: evolution,  
stars:Hertzsprung-Russell (HR) diagram, stars:horizontal branch 
\end{keywords}

\section{Introduction}  

The comparison of theoretical isochrones with observed Color Magnitude (CM) diagrams of
stellar clusters is the most direct method for testing evolutionary
theories and for investigating the evolutionary status of cluster
stars. In this context, the nearest open clusters have often been
selected in the past as privileged targets for testing theoretical
predictions. As early as 1963  Iben compared theoretical models with the Hyades and
Pleiades CM diagrams and with current estimates of the
luminosity-mass relationship. Though written about 40 years
ago, certain aspects of the paper are still of interest. The author regarded
the overlap of the Hyades and Pleiades Main Sequence (MS) as a risky procedure, 
due to the possibly different chemical compositions of the two clusters; 
a point that is still debated in recent papers, as we will discuss later. 
Moreover he drew attention to the role of superadiabatic convection, noting that 
there are no reasons for assuming the same mixing length value for stars of different chemical
composition, mass or in different evolutionary phases.

At that time, a main source of uncertainty in the comparison between
theory and observation was the unknown cluster distance, i.e. the
lack of observational constraints on the absolute magnitudes of the stars.
Hipparcos trigonometric parallaxes for the members of some nearby open
clusters has greatly improved the situation, allowing one to apply more stringent 
constraints to the theoretical predictions.  In a previous paper
(Castellani, Degl'Innocenti \& Prada Moroni 2001, Paper I) we compared
theoretical isochrones with improved Hipparcos observational data for
the Hyades (Dravins et al. 1997, Lindegren et al. 2000), discussing
the sources of uncertainties in the theoretical predictions.  More
recently, Madsen, Dravins \& Lindegren (2002) again used radial
motions of stars to present CM diagrams of unprecedented accuracy not
only for the Hyades but also for other nearby clusters, allowing further
useful comparisons with the theoretical scenario.  In this work we
have selected clusters with low or even negligible reddening estimates,
namely the Hyades, Pleiades and Ursa Major, in order to make theoretical
predictions on H burning structures at different stellar
metallicities.  We adopt the parallax of the single stars by Madsen et
al. (2002) (see their table 2). Observational data for Praesepe, whose
estimated metallicity is the same as that of the Hyades, will be added in
Sect.4 to discuss the observational evidence for the helium burning
evolutionary phase.

\section{Main sequence stars}


\begin{figure}
\label{castfig1}
\centerline{\epsfxsize=  8 cm \epsfbox{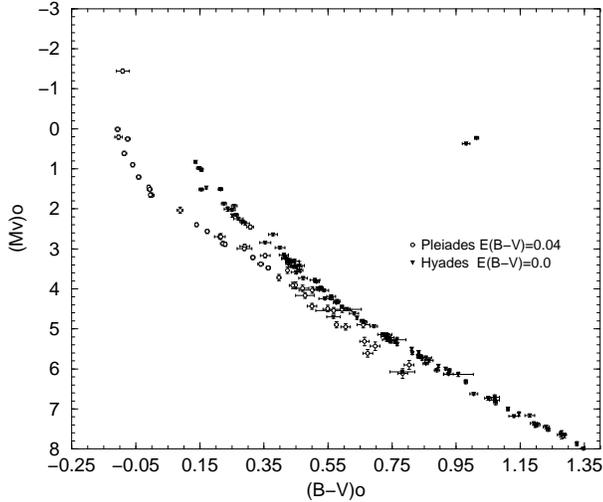}}
\caption{CMD for the Hyades and Pleiades clusters, using the parallax
values from Madsen et al. (2002). Visual, spectroscopical and
suspected binaries are excluded, see also Madsen et al. (2000) for the
Hyades and Raboud \& Mermilliod (1998) for the Pleiades. The Pleiades
data are corrected for the reddening. Error bars indicate observational errors as
given by Madsen et al. (2002) for the parallax and by the Hipparcos
catalog (at the node http://astro.estec.esa.nl/Hipparcos/HIPcataloguesearch.html) for the
colors.}
\end{figure}

\subsection{Hyades and Pleiades}

To start our investigation let us first consider the Hyades and Pleiades
as clusters having tighter MS, thereby imposing us more severe constraints
on the theoretical predictions.
 
Perryman et al. (1998) first used Hipparcos data to investigate
the distance, structure, membership, dynamics and age of the Hyades.
In Paper I we adopted Hyades parallaxes as improved by
Madsen et al. (2000) according to their kinematical method. In the
present paper we used the latest values given by Madsen et
al. (2002); however the differences in the Hyades CM diagram appear to be 
negligible. Regarding the Pleiades, Madsen et al. (2002) already observed that 
the kinematics does not improve the Hipparcos parallaxes. 

The reddening of the Hyades is generally assumed to be negligible (see
e.g. Perryman et al. 1998).  As well known, the interstellar
medium inside the Pleiades is not homogeneous, thus the cluster is
affected by a differential reddening (van Leeuwen 1983, Breger 1986,
Hansen-Ruiz \& van Leeuwen 1997) which produces a MS slightly scattered. 
As a first approximation we
adopted the commonly accepted average value E(B-V)$\approx$0.04 mag
(Robichon et al. 1999a, Mermilliod et al. 1997, van Leeuwen
1999a,Pinsonneault et al. 1998, Loktin, Matkin \& Gerasimenko 1994
etc.).

By adopting the above quoted values of the reddening and the distance moduli
we have drown the (Mv)$_{\mathrm o}$, (B-V)$_{\mathrm o}$ CM diagrams for
the two clusters shown in Fig.1; their relative location indicates
that the Pleiades stars should have lower metallicity and/or larger helium
content.

The aim of this work is to check if suitable theoretical models are able
to  reproduce the Hipparcos CM diagram for stars of
different metallicities. We are mainly interested in the region not
affected by external convection, that is a region in which the fit
does not depend on the free ``mixing length'' parameter. As already discussed in
Paper I, theoretical temperatures are indeed independent of the
efficiency of the external convection only for stars hotter than
B-V$\sim$0.4 (where the convection vanishes) or cooler than B-V$\sim$1.2
(where the convection becomes adiabatic). 
 
As in Paper I, we assume for the Hyades stars Z=0.024 (see e.g. Perryman et
al. 1998) together with Y=0.278, as given by extrapolation of the
linear relation between Y and Z, connecting metal poor Pop.II stars
(Z$={10}^{-4}$ Y$=0.23$) to the original composition of the Sun given by
standard solar models (SSM) Z$=0.02$  Y$=0.27$ (see e.g. Pagel \& Portinari 
1998, Castellani, Degl'Innocenti \& Marconi 1999). 

As for the Pleiades, recent estimates (Thevenin 1998, Friel \&
Boesgaard 1990, Grenon 1999), give -0.19 $\la$ [Fe/H] $\la$ 0.03. 
 It should be noted 
that [Fe/H]=0 does not necessarily correspond to the solar 
metallicity because the [Fe/H] value also depends on the helium content.
 Again according to SSM, the present metallicity and helium
abundance at the solar surface, after diffusion processes, are
estimated to be Z$\approx$0.017$\div$0.018 Y$\approx$0.24, reproducing
the observational value: (Z/X)$_{\odot} \sim 0.0230$ (see e.g. Bahcall,
Pinsonneault \& Basu 2001, Brun, Turck-Chi\`eze \& Zahn 1999, 
Ciacio, Degl'Innocenti \& Ricci 1997, Degl'Innocenti et al. 1997). 
Thus a star with present surface abundance of Z$\approx$0.02 Y$\approx$0.27
shows a value of [Fe/H] of about 0.06. As discussed below we will
adopt  the value Z=0.012 for the cluster fit,
which is within the observed range of metallicities.

Stellar models were computed with a  version of the
FRANEC evolutionary code (Chieffi \& Straniero 1989; Ciacio et al. 1997),
 improved so as to account for the most 
recent input physics (Cassisi et al. 1998), adopting OPAL EOS (Rogers et al. 1996) 
and using the Castelli (1999, C97) model atmospheres to derive stellar 
magnitudes in the selected photometric bands (see also Castelli 1998, 
Castelli, Gratton \& Kurucz 1997).


\begin{figure}   
\label{castfig2}
\centerline{\epsfxsize=  8 cm \epsfbox{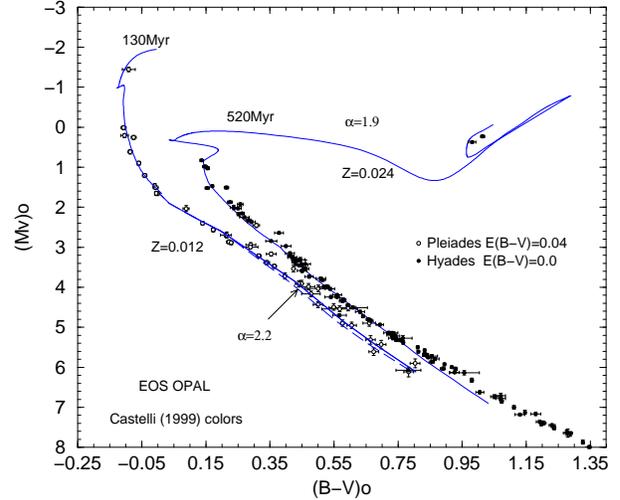}}
\caption{CM diagram of the Hyades and Pleiades, as in Fig.1, compared with
the theoretical isochrones for the Hyades and Pleiades composition
(Z=0.024 Y=0.278 and Z=0.012 Y=0.27, respectively).  The adopted
value for the mixing length is $\alpha$=1.9 in both cases. The
position of the Pleiades Zero Age Main Sequence (ZAMS) for a different
value of $\alpha$ is shown as dashed line. Equation of state from
Livermore (Rogers et al. 1996) and colour transformations and
bolometric corrections from Castelli (1999).}
\end{figure}

As well known, a decrease of metallicity or an increase of helium shifts the MS
toward higher temperatures. Thus to fit observational data one could
tune these two values to within reasonable ranges. If one fixes the metallicity at
Z=0.015, to fit observations one would need a helium abundance of
Y$\approx$0.30, a value which appears slightly too large.
Figure 2 shows our preferred  fit for the two clusters, as obtained 
for a Pleiades metallicity of Z=0.012, well within the range of
metallicity estimates, with a helium content of Y=0.27.  One finds that
theoretical results nicely reproduce observations in all regions excepting 
the lower end of the Hyades MS; this problem has already been analyzed
in Paper I and it will be not discussed here. Interestingly enough, one finds 
that the portion of the MS affected by external convection can be 
satisfactorly fitted with the same value of the mixing length parameter ($\alpha$=1.9). 

We conclude that the adopted  stellar models seem to be able to account for the
location of H burning structures even with different metallicities, and
that there is no evidence against the adopted evolutionary scenario.
The estimated age for the Pleiades appears to be in reasonable agreement with the results 
of Stauffer et al. (1999) and Ventura et al. (1998); however the 
range of evolutionary parameters derived in the literature for a given cluster, 
is an indication of the uncertainty still affecting  this kind of procedure.

\begin{figure}
\label{castfig3}
\centerline{\epsfxsize=  8 cm \epsfbox{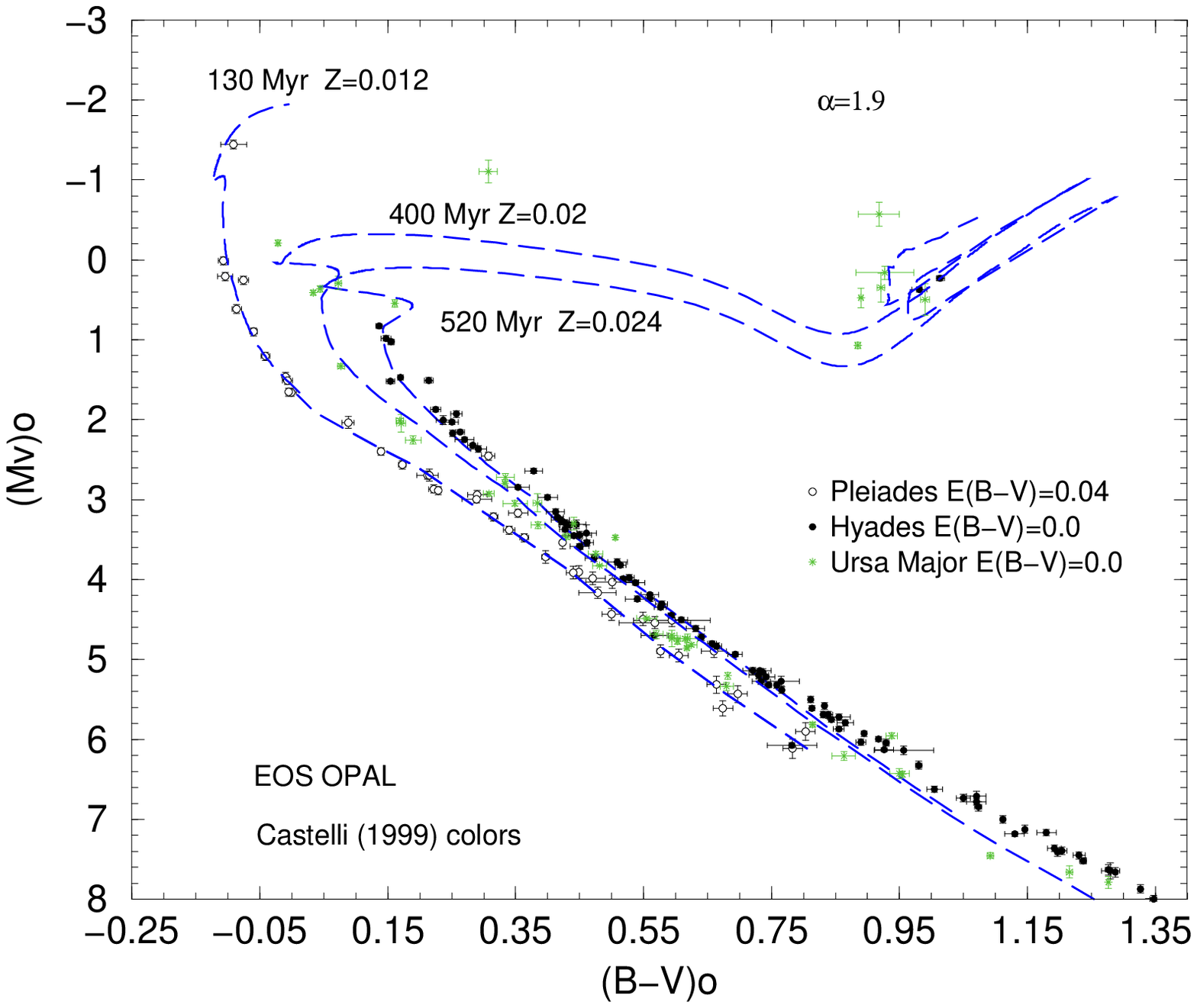}}
\centerline{\epsfxsize=  8 cm \epsfbox{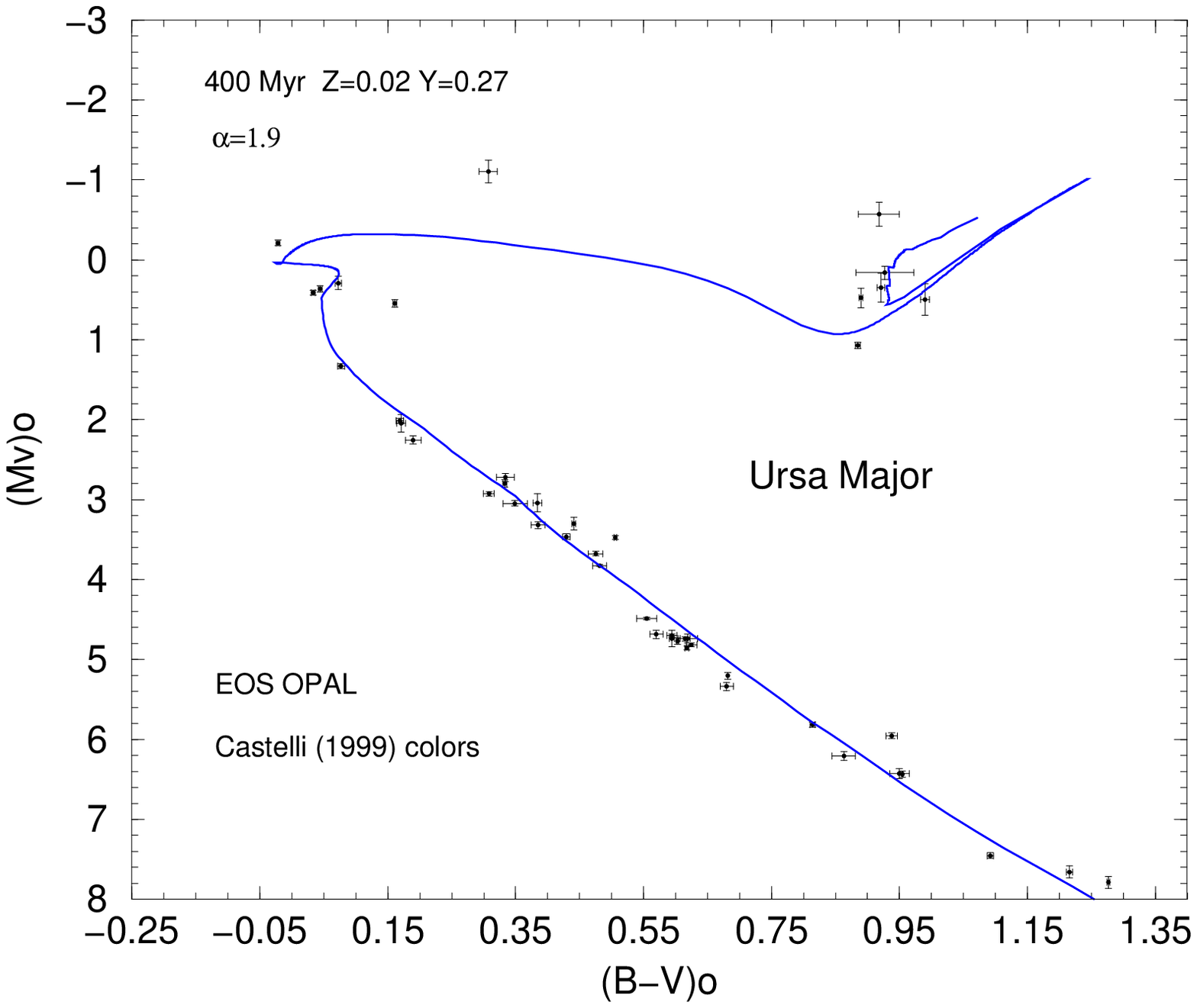}}
\caption{Upper panel: as in figure 2 with the addition 
of the fit of the  Ursa Major cluster. 
Lower panel: CMD for the  Ursa Major cluster, using the parallax
values from Madsen et al. (2002). Visual, spectroscopical and
suspected binaries are excluded. Error bars indicate observational errors as
given by Madsen et al. (2002) for the parallax and by the Hipparcos
catalog for the colors. Present best fit with a 400 Myr theoretical 
isochrone (Z=0.02 Y=0.27 $\alpha=1.9$) is also shown.
}
\end{figure}

\subsection{Ursa Major}

The previous investigation can be usefully implemented with data
for the Ursa Major, a cluster which - according to the literature -
should have a solar metallicity (Z=0.02) and a negligible reddening (Mermilliod 1977).  
Adopting  the photometric data from the Hipparcos catalogue and 
the trigonometric parallax from the radial velocities given by Madsen et al. (2002), one indeed finds 
(Fig.3, upper panel)  that the Ursa Major MS falls just between the Hyades 
and Pleiades MS, confirming the quoted metallicity.

Figure 3, lower panel, shows our best fit for the Ursa Major cluster 
as obtained for Z=0.02, Y=0.27 and an age of 400 Myr. 
The agreement between the observed data and the theoretical isochrones appears
 to be satisfactory even if the statistic is poor.  
It may be noted that the phases influenced by the $\alpha$ value appear to be fitted by 
the same value of the mixing-length parameter ($\alpha=1.9$) 
for all the three clusters.

Before leaving the argument, it is worth noting that present models have 
all been computed by adopting the classical Schwarzschild 
criterion for the extension of the convective regions. As well known, the
debate is still open about the occurrence of a larger central mixing
as given by the core overshooting scenario (see, e.g., Testa et al. 1999, Pols et al. 1998). 
However, as already shown for the Hyades in Paper I, 
we found that a similar best fit for all the clusters
can be attained with mild overshooting (extra-mixing by $l_{ov}=0.25 H_p$)
provided that the cluster age is increased (see e.g. Maeder 1976).
The reason is that the two sets of isochrones significantly differ only in 
the depopulated region of the CM diagram. 

\section{Helium burning stars}
\begin{figure}
\label{castfig4}
\centerline{\epsfxsize=  8 cm \epsfbox{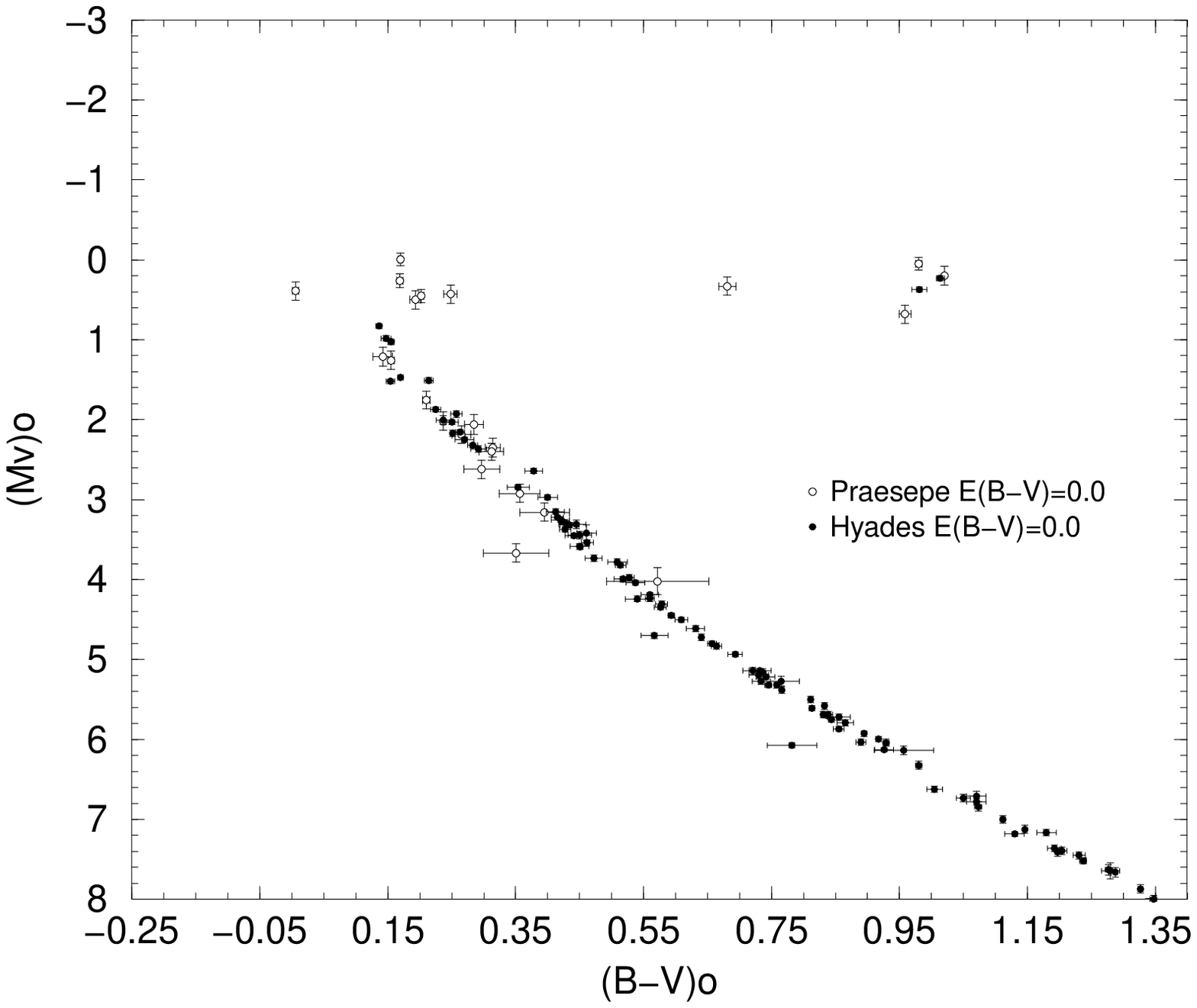}}
\centerline{\epsfxsize=  8 cm \epsfbox{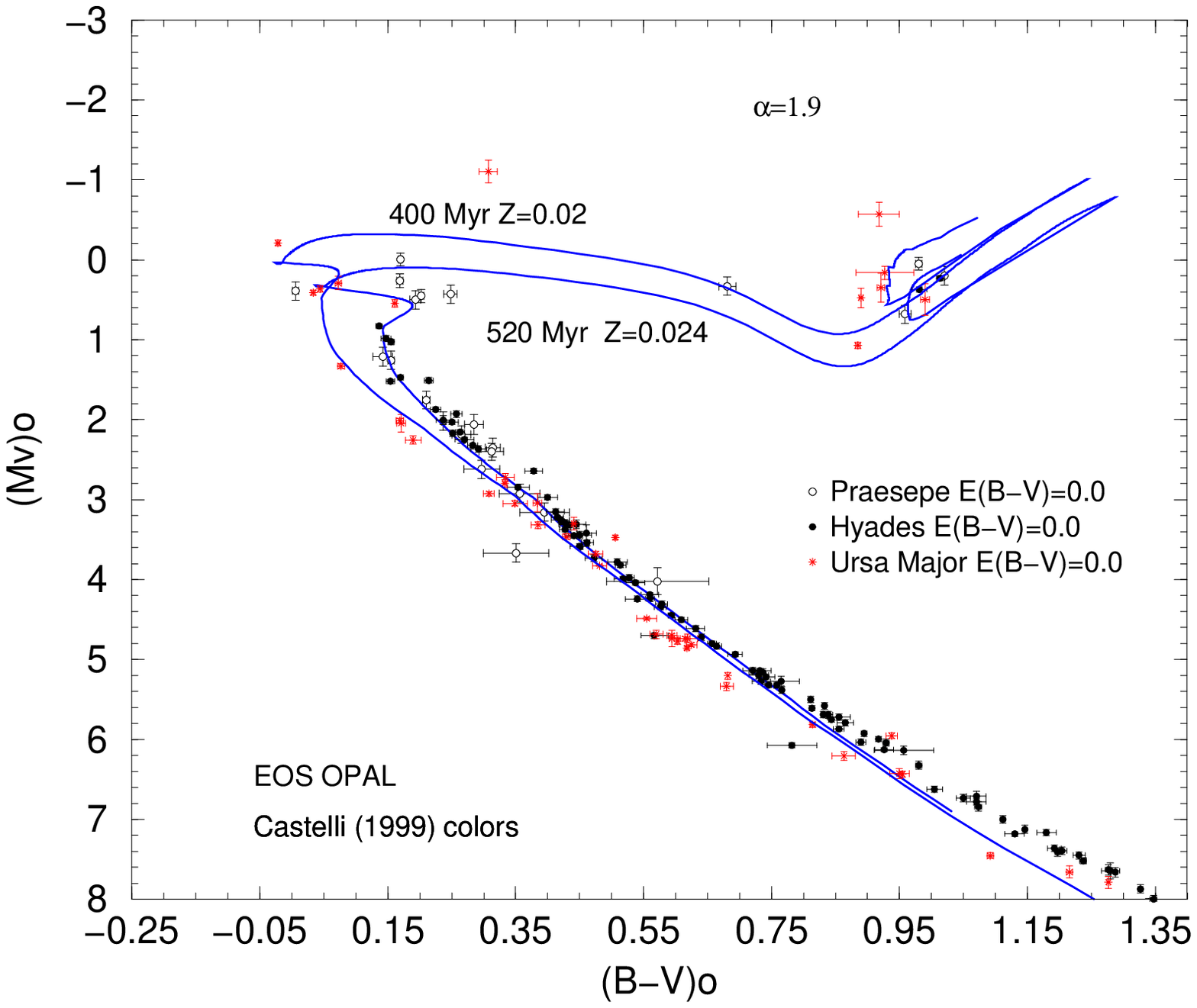}}
\caption{Upper panel: CMDs for the  Hyades and the Praesepe clusters, using the parallax
values from Madsen et al. (2002). Visual, spectroscopical and
suspected binaries are excluded. Error bars indicate observational errors as
given by Madsen et al. (2002) for the parallax and by the Hipparcos
catalog for the colors. Lower panel: CMDs for the  Hyades, Praesepe and Ursa Major clusters and 
the relating isochrones.
}
\end{figure}
 
As shown in the previous Figures 2 and 3, both the  Hyades and  Ursa Major have 
He-burning stars, though not very abundant. In Paper I we discussed 
He-burning stars in the Hyades, advancing the suggestion for theoretical models 
slightly underluminous.  This is a relevant point since for less massive stars 
with degenerate progenitors an opposite tendency has been suggested (see e.g. Pols et
al. 1998; Castellani et al. 2000). 

However, the sample of Hyades He-clump stars can be improved by adding 
the data from  Praesepe. Indeed metallicity estimates for Praesepe 
provide values which are compatible with the Hyades metallicity (see e.g. $[Fe/H]=0.135\pm 0.07$
by Boesgaard \& Budge 1988, $[Fe/H]=0.17\pm 0.01$ by Grenon 1999). 
Such a similarity is supported by the similar CM diagram location of the two
MS, as shown in Figure 4 where Praesepe data are plotted taking the
photometry from the Hipparcos catalogue, the trigonometric
parallax from Madsen et al. (2002) and assuming a negligible reddening 
(Mermilliod et al. 1997, Robichon et al. 1999a, Pinsonneault et al. 1998, van Leeuwen 1999a).

Figure 4 also shows that the two clusters have quite similar star distributions
 all over the CM diagram, supporting not only a similar metallicity 
but also a similar age, allowing one to adopt the He-burning stars as a unique common sample.
The number of stars in the He-burning clump remains rather small,
not allowing us to make precise constraints on the theoretical
models. However, with this caveat in mind, Fig. 4
(lower panel) now seems to suggest a general agreement between the theory
and the observational data, casting serious doubts on the suspected underluminosity
of the models.  

 Interestingly enough, one also finds that the Ursa Major He-clump 
not only appears to be in fair agreement with the theoretical predictions but 
 that it also appears to be brighter (on the average) than those of the Hyades 
and Praesepe, as predicted by theory as a consequence of the lower age.  
Unfortunately the He-clump is quite spread out, so
that the observational constraints are not as stringent as
one would wish. The only conclusion that can be safely drawn is that there is a general
 agreement between the theory and observation even though further investigations
are needed. Incidentally, we recall that the situation is
 not changed by the adoption of a mild core overshooting during the main
 sequence phase (see e.g. Paper I for a discussion).

\section{Discussion and conclusions}

To conclude the paper it may be worth recalling briefly
the debate concerning the Pleiades CM diagram.
In recent years, several authors, assuming for the Pleiades a solar metallicity (see e.g
 Mermilliod et al. 1997, Pinsonneault et al. 1998), found
a disagreement between the models and the Hipparcos data. To account for such
a disagreement,  Pinsonneault et al. (1998) suggested the possibility of 
localized systematic errors in the Hipparcos parallaxes of the order 
of 1 mas for open clusters and  stellar associations. The disagreement 
was quite recently confirmed by Stello \& Nissen (2001) on the basis of Str\"omgren
 photometry of F-type stars.  

Narayanan \& Gould (1999) used Hipparcos proper motions to
further investigate the parallaxes of Pleiades stars, finding a 
distance modulus with a  rather large error ($\pm$ 0.18 mag.)  which they claimed to be 
in disagreement with that  derived directly from Hipparcos parallaxes 
and in agreement with that  obtained through MS fitting.
However, the uncertainty is of the same order as that of the quoted discrepancy.  
Thus the authors  suggested the possibility of systematic errors due to spatial
correlations over small angular scales. 

This possibility was rejected  by Robichon et al. (1999a,b) and van Leeuwen (1999b) 
who, on the basis of a range of statistical checks on the data and an evaluation of data reduction methods,
 excluded the occurrence of  systematic errors (see also van Leeuwen \& Evans 1998).  Moreover Stello \& Nissen (2001) 
 quoted the  recent photometric determination of the Pleiades 
metallicity by  Grenon (1999) based on about 62 stars: [Fe/H]=-0.11$\pm$0.025
showing that with this choice of metallicity the Pleiades ZAMS is
fitted successfully, as supported and confirmed by the present investigation. 

We conclude that, if the metallicities adopted in this work
 will be confirmed by future
investigations, theoretical models appear to be consistent 
with observations and there are no reasons for claiming the existence of errors 
either in the  Hipparcos results or in the theoretical predictions. 
At the present status of the art,  uncertainties in the chemical composition of 
the Pleiades are larger than uncertainties in  the Hipparcos parallaxes and thus, 
in our opinion,  it is more reasonable to search for the Pleiades composition within plausible ranges
of metallicities, as we did, rather than to follow the
opposite procedure (see Lebreton 2001 for a discussion).

\section{Acknowledgements}

We warmly thank D. Dravins, L. Lindegren and S. Madsen for providing
us improved Hipparcos data before publication and the referee Floor van Leeuwen for 
useful comments. This work is partially
supported by the Ministero dell'Universit\`a della Ricerca Scientifica
(MURST) within the ``Stellar observables of cosmological relevance''
project (COFIN2000).

\label{lastpage}


\begin{thebibliography}{99}

\bibitem{} Bahcall J.N., Pinsonneault M.H., Basu S., 2001, ApJ, 555, 990 
\bibitem{} Boesgaard A.M., Budge K.G., 1988, ApJ, 332, 410
\bibitem{} Breger, M., 1986, ApJ 309, 311
\bibitem{} Brun A.S., Turck-Chi\`eze S., Zahn J.P., 1999, ApJ, 525, 1032
\bibitem{} Cassisi S., Castellani V., Degl'Innocenti S., Weiss A., 1998, A\&A Suppl., 129, 267
\bibitem{} Castellani V., Degl'Innocenti S., Marconi M., 1999, A\&A, 349, 834
\bibitem{} Castellani V., Degl'Innocenti S., Girardi L., Marconi M., Prada Moroni P.G., Weiss A., 2000, A\&A, 354, 150
\bibitem{} Castellani V., Degl'Innocenti S., Prada Moroni P.G., 2001, MNRAS, 320, 66 (Paper I)  
\bibitem{} Castelli F., 1998, Memorie della Societa Astronomica Italiana, 69, 165 
\bibitem{} Castelli F., 1999, A\&A, 346, 564
\bibitem{} Castelli F., Gratton R. G., Kurucz R. L., 1997, A\&A, 318, 841
\bibitem{} Chieffi A., Straniero O., 1989, ApJS, 71, 47
\bibitem{} Ciacio F., Degl'Innocenti S., Ricci B., 1997, A\&AS, 123, 449   
\bibitem{} Degl'Innocenti S., Dziembowski W.A., Fiorentini G., Ricci B., 1997, Astrop. Phys., 7, 77
\bibitem{} Dravins D., Lindegren L., Madsen S., Holmberg J., 1997, ESA, SP-402, p. 733
\bibitem{} Friel E., Boesgaard A.M., 1990, ApJ, 387, 480
\bibitem{} Grenon M., 1999, in Proc. of 11th cambridge workshop on cool stars, in press 
\bibitem{} Hansen-Ruiz C.S., van Leeuwen F., in ``Hipparcos-Venice '97'', ESA SP-402, pag. 295 
\bibitem{} Iben I., 1963, ApJ, 138, 452
\bibitem{} Lebreton Y., 2001, ARA\&A, 38, 35
\bibitem{} Lindegren L., Madsen S., Dravins D., 2000, A\&A, 356, 1119
\bibitem{} Loktin A.V., Matkin N.V., Gerasimenko T.P., 1994, A\&AT, 4, 153
\bibitem{} Maeder A., 1976, A\&A, 47, 389
\bibitem{} Madsen S., Lindegren L., Dravins D., 2000, in ASP Conf. Ser. 198, Stellar Cluster and Associations, ed. R. Pallavicini, G. Micela, S. Sciortino, 137
\bibitem{} Madsen S., Dravins D., Lindegren L., 2002, A\&A, 381, 446
\bibitem{} Mermilliod J.C., 1977, Bull. Inf. Centre Donnees Stellaires, Vol. 12, page 2
\bibitem{} Mermilliod J.C., Turon C., Robichon N., Arenou F., Lebreton Y., 1997, in ``Hipparcos-Venice '97'', ESA SP-402, page 643
\bibitem{} Narayanan V. K., Gould A., 1999, ApJ, 523, 328
\bibitem{} Pagel B.E.J., Portinari L., 1998, MNRAS, 298, 747  
\bibitem{} Perryman M.A.C. et al., 1998, A\&A, 331, 81   
\bibitem{} Pinsonneault M.H., Stauffer J., Soderblom D.R., King J.R., Hanson R.B., 1998, ApJ, 504, 170
\bibitem{} Pols O.R., Schroeder K-P, Hurley J.R., Tout C.A., Eggleton P.P., 1998, MNRAS, 298, 525
\bibitem{} Raboud D., Mermilliod J.C., 1998, A\&A, 329, 101 
\bibitem{} Robichon N., Arenou F., Lebreton Y., Turon C., Mermilliod J.C., 1999a, in ``Harmonizing cosmic distances and scales in a post Hipparcos Era'', ASP Conference Series, 167, 72, Egret and Heck editors
\bibitem{} Robichon N., Arenou F., Mermilliod J.C., Turon C., 1999b, A\&A, 345, 471
\bibitem{} Rogers F. J., Swenson F. J., Iglesias C. A., 1996, ApJ, 456, 902 
\bibitem{} Stauffer J.R. et al., 1999, ApJ, 527, 219
\bibitem{} Stello D., Nissen P.E., 2001, A\&A 374, 105
\bibitem{} Testa V., Ferraro R., Chieffi A., Straniero O., Limongi M., Fusi Pecci F., 1999, AJ, 118, 2839
\bibitem{} Thevenin F., 1998, Chemical abundances in late-type stars, available on the web at the node http://vizier.u-strasbg.fr
\bibitem{} van Leeuwen F., 1983, PhD Thesis, Leiden University 
\bibitem{} van Leeuwen F., 1999a, in ``Harmonizing cosmic distances and scales in a post Hipparcos Era'', ASP Conference Series, 176, 52, Egret and Heck editors
\bibitem{} van Leeuwen, F. , 1999b, A\&A, 341, L71  
\bibitem{} van Leeuwen, F., Evans D.W., 1998, A\&AS 130, 157
\bibitem{} Ventura P., Zeppieri A., Mazzitelli I., D'Antona F., 1998, A\&A 334, 953 
\end{thebibliography}
\end{document}